\pdfoutput=1
\documentclass[aps, prx, reprint, superscriptaddress, showkeys]{revtex4-2}

\usepackage[utf8]{inputenc}
\usepackage{graphicx}
\usepackage{amsmath}
\usepackage{amssymb}
\usepackage[colorlinks=true, linkcolor=blue, citecolor=blue, urlcolor=blue]{hyperref}

\begin{document}
	
	\title{\Large In-Plane Differentiation and Out-of-Plane Coherence in Superconducting K$_{2-x}$Fe$_{4+y}$Se$_5$}
	
	\author{Chih-Han Wang}
	\affiliation{Institute of Physics, Academia Sinica, Taipei 11529, Taiwan}
	
	\author{Wu Phillip M}
	\affiliation{Institute of Physics, Academia Sinica, Taipei 11529, Taiwan}
	
	\author{Gwo-Tzong Huang}
	\affiliation{Institute of Physics, Academia Sinica, Taipei 11529, Taiwan}
	
	\author{Ming-Jye Wang}
	\affiliation{Institute of Astrophysics and Geometrical Astronomy, Academia Sinica, Taipei 10617, Taiwan}
	
	\author{Maw-Kuen Wu}
	\email{mkwu@phys.sinica.edu.tw}
	\affiliation{Institute of Physics, Academia Sinica, Taipei 11529, Taiwan}
	
	\date{\today}
	
\begin{abstract}
	The structural link between superconductivity and anomalous $d_2$ diffraction signatures in K$_{2-x}$Fe$_{4+y}$Se$_5$ remains unresolved. Their appearance is accompanied by pronounced host-lattice evolution rather than a static structural background. Using high-resolution synchrotron X-ray diffraction, we show that excess Fe alters the vacancy-ordered host lattice, with structural differences between superconducting and nonsuperconducting states emerging before the $d_2$ features appear. All observed $d_2$ reflections can be described using two characteristic in-plane spacings and a nearly common low-temperature $c$ parameter. Temperature-dependent measurements reveal that superconductivity is associated with persistent in-plane differentiation and enhanced Fe-layer coherence along the $c$ axis. These results establish host-lattice evolution as the common structural origin linking the $d_2$ family and superconductivity, and suggest that the host-lattice differentiation identified here represents a structural analogue of the symmetry redistribution in FeSe.
\end{abstract}

	\maketitle
	
\section{Introduction}

    Understanding how lattice structure governs superconductivity remains one of the central challenges in condensed-matter physics \cite{1,2,3,4,5,6,7,8}. In several major families of high-temperature superconductors, superconductivity emerges together with a redistribution of structural symmetry within a contiguous crystallographic framework. A representative example is the cuprate YBa$_2$Cu$_3$O$_{7-\delta}$ (YBCO), where superconductivity develops within an orthorhombic lattice characterized by inequivalent in-plane lattice parameters ($a \neq b$). A remarkably similar structural motif is found in FeSe, where the superconducting state is accompanied by a tetragonal-to-orthorhombic transition and the emergence of inequivalent Fe--Fe environments within the Fe plane \cite{9,10,11,12,13,14}. Despite their different electronic structures, both systems share an important structural characteristic: superconductivity develops within a continuous lattice that supports differentiated in-plane environments while preserving a common structural framework along the out-of-plane direction.
    
    In contrast, the alkali-metal-intercalated iron selenide K$_{2-x}$Fe$_{4+y}$Se$_5$ presents a unique structural challenge. Although it is constructed from the same FeSe building blocks, superconductivity emerges from a vacancy-ordered antiferromagnetic parent phase rather than from a metallic state. The parent K$_2$Fe$_4$Se$_5$ compound exhibits long-range Fe-vacancy ordering and is generally regarded as an insulating antiferromagnetic ground state \cite{15,16,17}. Superconductivity appears only after excess Fe occupies vacancy-related sites, making Fe occupation of vacancy-related sites one of the most important parameters controlling the structural and electronic evolution of this material family \cite{18,19,20,21,22}. Unlike YBCO and FeSe, however, the superconducting state of K$_{2-x}$Fe$_{4+y}$Se$_5$ has frequently been interpreted within a phase-separation framework involving structurally distinct components with independent lattice parameters \cite{23,24,25,26,27,28,29,30,31,32}. Within such descriptions, the additional diffraction signatures are commonly associated with a second structural environment possessing not only distinct in-plane lattice spacings but also an independent out-of-plane lattice constant.
    
   Interestingly, scanning tunneling microscopy studies have revealed local electronic inhomogeneities and region coexistence in K-doped iron selenides, suggesting that superconductivity is closely tied to the evolution of the host lattice rather than an isolated minority phase \cite{21}. Traditionally, however, the additional diffraction features have generally been viewed as evidence for structurally separate components rather than structural differentiation developing within a common crystallographic framework.
    
    Previous synchrotron diffraction studies demonstrated that excess Fe strongly modifies the structural response of the vacancy-ordered host lattice and that superconductivity is closely associated with the suppression of the large lattice rearrangement observed in the parent compound \cite{19}. These results established excess Fe as a key structural control parameter and revealed that superconducting and nonsuperconducting samples follow fundamentally different structural evolution pathways despite retaining similar average crystallographic symmetry. Importantly, the structural divergence between the two states develops within the dominant host lattice itself, suggesting that the evolution of the host framework may be more relevant to superconductivity than the appearance of additional diffraction features alone. However, the microscopic structural pathway linking excess-Fe incorporation, host-lattice evolution, and superconductivity remains unresolved.
    
    At the same time, studies of structurally complex materials have demonstrated that substantial local structural differentiation can coexist within a common crystallographic framework and generate additional diffraction signatures without requiring complete macroscopic phase separation \cite{33,34,35,36,37,38}. An important unresolved question therefore concerns whether the additional diffraction features indeed require multiple crystallographic components, or whether they instead arise from structural differentiation within a single evolving host lattice.
    
    In this work, we investigate the structural evolution of K$_{2-x}$Fe$_{4+y}$Se$_5$ as a function of Fe content, temperature, and thermal history using high-resolution synchrotron X-ray diffraction. We identify two characteristic structural features of the superconducting state: in-plane structural differentiation and enhanced coherence between Fe layers along the $c$ axis. By directly comparing superconducting and nonsuperconducting compositions, we show that these characteristics develop prior to the full appearance of the additional diffraction signatures. We further demonstrate that the complete experimentally observable diffraction family can be understood within a host-lattice framework characterized by two distinct in-plane spacings and a nearly common low-temperature $c$-axis parameter. The resulting picture establishes in-plane differentiation and out-of-plane coherence as the key structural characteristics linking host-lattice evolution, the additional diffraction signatures, and superconductivity in K$_{2-x}$Fe$_{4+y}$Se$_5$.

\section{Experimental Procedures}
    The $\mathrm{K_2Fe_4Se_5}$ (parent phase) and superconducting $\mathrm{K_{1.9}Fe_{4.2}Se_5}$ polycrystalline samples were synthesized using a solid-state reaction method developed in our previous work \cite{39} and further optimized to ensure chemical homogeneity and control the final structural state. High-purity Fe, Se, and K metal plates were mixed according to the target stoichiometry ($\sim 5$~g batches). To ensure uniform precursor mixing, the reagents were subjected to high-energy mechanical ball milling for 60 minutes, a process repeated three times. Aliquots of the resulting powder ($\sim 0.5$~g) were sealed in quartz ampoules backfilled with high-purity argon.

    The thermal history was strictly controlled to isolate the distinct phases: (1) For the parent phase ($\mathrm{K_2Fe_4Se_5}$), samples were slowly heated to 650$^\circ\mathrm{C}$--840$^\circ\mathrm{C}$ over 7 hours, annealed for 24--30 hours, and subsequently furnace-cooled to room temperature. This slow cooling path ensures the formation of the thermodynamically stable, vacancy-ordered ground state. (2) For the superconducting phase ($\mathrm{K_{1.9}Fe_{4.2}Se_5}$), samples were inserted directly into a pre-heated furnace set to the desired quenching temperature ($T_Q$, 650$^\circ\mathrm{C}$--840$^\circ\mathrm{C}$), held for 2 hours for equilibration, and then rapidly quenched into ice water. This fast-cooling protocol is critical for preserving the metastable structural state associated with excess-Fe incorporation, which is further analyzed below in relation to lattice matching and superconductivity.

    High-resolution synchrotron X-ray diffraction (XRD) data were collected at the National Synchrotron Radiation Research Center (NSRRC), Taiwan. Magnetization measurements were performed using a SQUID magnetometer (Quantum Design). The structural refinements were carried out using the Rietveld method.
	
	\section{Results and Discussion}
	
	\begin{figure*}[t]
		\centering
		\includegraphics[width=0.9\textwidth]{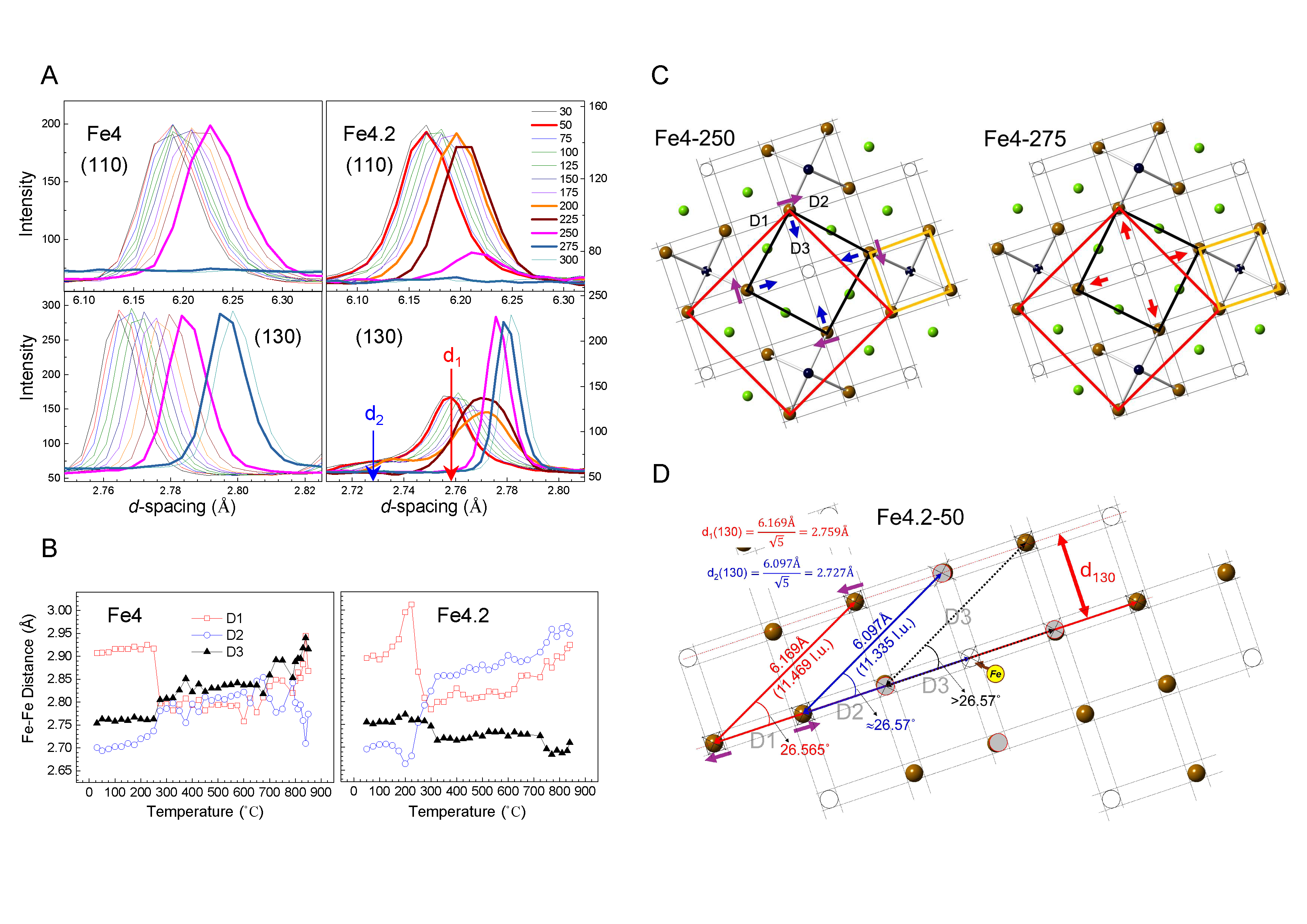}
\caption{\textbf{Host-lattice evolution and the origin of the additional diffraction features.}
	(A) Temperature-dependent evolution of the (110) and (130) reflections for Fe$_4$ and Fe$_{4.2}$. The Fe$_{4.2}$ sample develops additional diffraction features (d$_2$) upon cooling. (B) Temperature dependence of the nearest-neighbor Fe--Fe distances obtained from refinement of the I4/m host lattice. D1, D2, and D3 arise from the geometrical relationship between Fe atoms and Fe-vacancy sites shown in (C). (C) Vacancy-ordering-induced structural rearrangement in Fe$_4$, producing selective bond contraction and rotational distortion of the Fe network. (D) Real-space reconstruction derived from the refined Fe$_{4.2}$ host lattice. Partial occupation of vacancy-related sites by excess Fe suppresses the vacancy-ordering distortion and generates differentiated local environments within a single host lattice. The projected spacings reproduce the observed d$_2$ positions. Here, l.u. denotes a length unit used in the geometric construction.}
		\label{fig:fig1}
	\end{figure*}
	
    To investigate the structural origin of the additional diffraction features observed in superconducting K$_{2-x}$Fe$_{4+y}$Se$_5$, we first examine the temperature-dependent diffraction evolution of the nonsuperconducting Fe$_4$ and superconducting Fe$_{4.2}$ samples (Fig.~\ref{fig:fig1}A). The (110) reflection corresponds to the characteristic Fe-vacancy superlattice peak, whereas the (130) reflection is associated with the nearest interlayer Fe-layer spacing. The Fe$_4$ sample exhibits a pronounced structural response across the vacancy-ordering temperature range, accompanied by a clear shift of the diffraction peaks that reflects substantial lattice contraction. In contrast, the Fe$_{4.2}$ sample exhibits a much weaker lattice response over the same temperature interval and eventually develops additional diffraction features (d$_2$) near the (130) reflection. The suppression of the lattice response immediately suggests that excess Fe alters the structural evolution associated with the vacancy-ordering process.
    
    To understand this behavior, we analyze the nearest-neighbor Fe--Fe distances obtained from refinement of the dominant host lattice (Fig.~\ref{fig:fig1}B). Following the conventional approach adopted in previous studies, the diffraction patterns were refined using an I4/m host structure together with an I4/mmm component associated with the additional diffraction features. Importantly, all structural parameters discussed below are derived exclusively from refinement of the I4/m host lattice.
    
    The three characteristic Fe--Fe distances, D1, D2, and D3, originate from the geometrical relationship between Fe atoms and Fe-vacancy sites illustrated in Fig.~\ref{fig:fig1}C. In the Fe$_4$ parent compound, D1, D2, and D3 progressively converge upon heating and become nearly identical in the high-temperature state, consistent with the recovery of the average I4/mmm structure. This behavior indicates that the Fe sublattice approaches a nearly equivalent atomic configuration at high temperature. In contrast, the Fe$_{4.2}$ sample maintains distinct D1, D2, and D3 even at high temperature, indicating that excess Fe preserves local structural differentiation within the host lattice and prevents complete convergence toward the average Fe-sublattice configuration.
    
    A particularly significant difference is observed in D3 during the structural transition. In Fe$_4$, D3 decreases markedly upon cooling, reflecting the lattice contraction associated with the rearrangement of the Fe sublattice. In contrast, this contraction is strongly suppressed in Fe$_{4.2}$, indicating that excess Fe inhibits the structural rearrangement that normally develops during cooling. The evolution of D1 further supports this picture. A pronounced change in D1 occurs near 250$^\circ$C and coincides with the growth of the (110) superlattice reflection in Fig.~\ref{fig:fig1}A, demonstrating a direct connection between D1 and the development of vacancy-related structural correlations. At the same time, D1 and D2 evolve in an approximately opposite manner, indicating that the host lattice develops increasingly differentiated local environments rather than undergoing a uniform structural distortion.
    
    Another notable feature is the appearance of extrema in D1, D2, and D3 near $\sim200^\circ$C. This temperature closely coincides with the emergence of the d$_2$ features. The strong correspondence between these extrema and the formation of d$_2$ demonstrates that the additional diffraction features are intimately connected to host-lattice evolution. More importantly, the structural divergence between $\mathrm{Fe}_4$ and $\mathrm{Fe}_{4.2}$ develops between approximately $275\,^\circ\mathrm{C}$ and $250\,^\circ\mathrm{C}$, while the $d_2$ features are still absent. This temporal sequence suggests that host-lattice differentiation precedes the formation of the additional diffraction features. Upon further cooling, a notable series of extrema appears in $D_1$, $D_2$, and $D_3$ near $\sim 200\,^\circ\mathrm{C}$ (as shown in Fig.~1B). This specific temperature precisely marks the critical onset where the $d_2$ features begin to emerge and germinate in the diffraction profiles. Connecting this sequence—starting with the initial host-lattice differentiation at $275\,^\circ\mathrm{C}-250\,^\circ\mathrm{C}$, followed by the structural adjustments culminating in extrema at $\sim 200\,^\circ\mathrm{C}$—demonstrates that the appearance of the $d_2$ diffraction signatures is a direct consequence of this evolving host-lattice differentiation rather than its initial trigger.
    
    The structural origin of this differentiated state is illustrated in Fig.~\ref{fig:fig1}C. In the Fe$_4$ parent compound, cooling drives a rearrangement of the Fe sublattice relative to the vacancy framework. As Fe atoms move toward their low-temperature positions, selective bond contraction is accompanied by a rotational rearrangement of the Fe network, lowering the average symmetry from the high-temperature I4/mmm configuration toward the vacancy-ordered I4/m structure. The resulting geometry naturally gives rise to the three characteristic distances D1, D2, and D3 observed in Fig.~\ref{fig:fig1}B.
    
    Based on this refined structural framework, we construct the model shown in Fig.~\ref{fig:fig1}D. It is important to emphasize that Fig.~\ref{fig:fig1}D is not a schematic illustration but a real-space reconstruction derived directly from the refined atomic positions of the Fe$_{4.2}$ host lattice. We consider the effect of excess Fe partially occupying vacancy-related sites, thereby locally suppressing the contraction and rotational rearrangement that normally accompany cooling. As a consequence, two distinct local structural environments naturally emerge within a single continuous host lattice: one retaining the full structural rearrangement and the other exhibiting a partially suppressed rearrangement.
    
    Remarkably, the projected spacings generated from these differentiated local environments reproduce the experimentally observed positions of the d$_2$ diffraction features. Since the model is constructed entirely from the refined host lattice and does not utilize structural information extracted from the d$_2$ component itself, this agreement demonstrates that the host lattice already contains the essential structural information required to account for the additional diffraction features.
    
    These results provide a direct real-space picture of how excess Fe modifies the host lattice. The suppression of the vacancy-ordering distortion generates differentiated local environments within the Fe plane and establishes a structural framework that precedes the appearance of the additional diffraction features. The emergence of the $d_2$ diffraction signatures therefore follows the development of host-lattice differentiation rather than initiating it. In this picture, the $d_2$ features represent diffraction manifestations of an already differentiated host lattice.
	
	It is particularly noteworthy that the characteristic lattice-parameter evolution across the structural transition region is not an artifact unique to our polycrystalline samples~\cite{19}, as crushed single-crystal neutron diffraction studies reveal analogous behaviors~\cite{40}. Furthermore, a careful examination of their superconducting sample reveals that the two c-axis lattice parameters associated with the superconducting variant progressively converge upon cooling. Moreover, their powder neutron diffraction analysis adopted an orthorhombic space group (Fmmm), rather than the conventional I4/mmm description, for the superconducting phase, yielding a finite in-plane anisotropy ($a \neq b$). Although interpreted within a phase-separation framework, these observations reveal structural characteristics that closely parallel the in-plane differentiation identified in the present work.
	\begin{figure*}[t]
		\centering
		\includegraphics[width=0.8\textwidth]{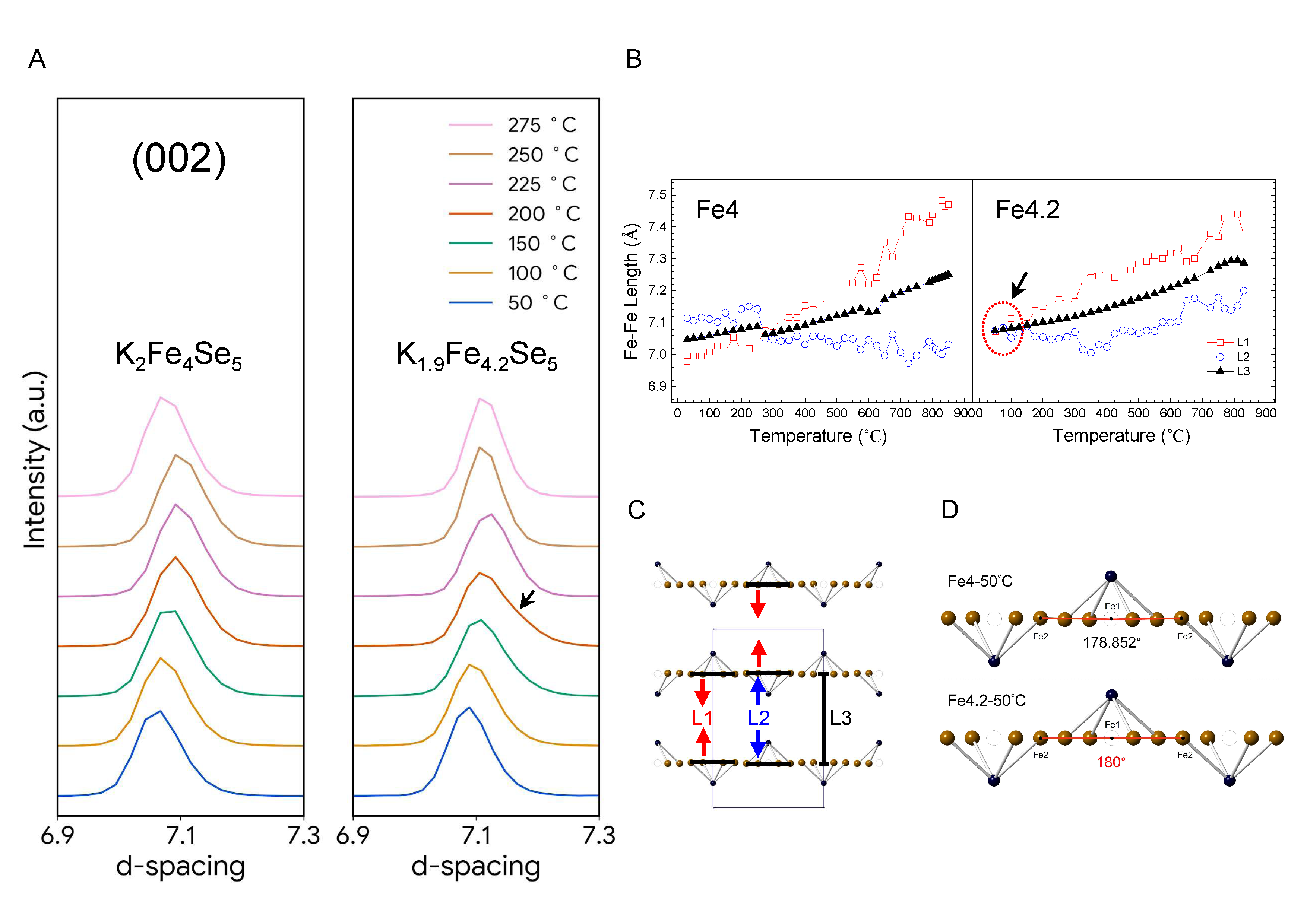}
		\caption{\textbf{Out-of-plane host-lattice evolution and Fe-layer geometry}
			(A) Temperature-dependent evolution of the (002) reflection for Fe$_4$ and Fe$_{4.2}$. A d$_2$ shoulder emerges near $\sim200^\circ$C in the Fe$_{4.2}$ sample. (B) Temperature dependence of the interlayer Fe--Fe distances L1, L2, and L3. L1 and L2 remain distinct in Fe$_4$ but progressively converge in Fe$_{4.2}$ upon cooling. (C) Geometric definition of the interlayer distances L1, L2, and L3. (D) Refined low-temperature structures. The Fe$_4$ lattice retains layer mismatch (L1 $\neq$ L2), whereas Fe$_{4.2}$ approaches L1 = L2, yielding a nearly ideal 180$^\circ$ Fe-layer configuration.}
		\label{fig:fig2}
	\end{figure*}
	
	We now turn to the out-of-plane structural evolution, which provides further insight into how excess Fe modifies the host lattice. Figure~\ref{fig:fig2}A shows the temperature dependence of the (002) reflection for Fe$_4$ and Fe$_{4.2}$ samples. In the Fe$_4$ parent compound, the (002) reflection exhibits a pronounced structural response between approximately 275$^\circ$C and 250$^\circ$C, indicating a substantial rearrangement of the host lattice during cooling. In contrast, the Fe$_{4.2}$ sample shows a much weaker evolution across the same temperature range. This behavior closely parallels the suppressed D3 contraction observed in Fig.~\ref{fig:fig1}B, demonstrating that excess Fe suppresses both the in-plane and out-of-plane structural responses of the host lattice.
	
	A particularly important feature appears near $\sim 200\,^\circ\mathrm{C}$ in the $\mathrm{Fe}_{4.2}$ sample. At this temperature, a shoulder develops on both the (130) and (002) reflections, corresponding to the emergence of the $d_2$ component along both in-plane and out-of-plane directions. This $\sim 200\,^\circ\mathrm{C}$ anomaly acts as a synchronized structural milestone: it coincides perfectly with the extrema observed in the Fe-Fe distances ($D_1, D_2, D_3$) and the interlayer parameters ($L_1, L_2$). This tight temporal correlation—bridging the early-stage host-lattice divergence at $275\,^\circ\mathrm{C}-250\,^\circ\mathrm{C}$ and the subsequent $d_2$ germination at $\sim 200\,^\circ\mathrm{C}$—underscores that the $d_2$ state represents a unified, global structural response of the evolving host lattice.
	
	The temperature dependence of L1, L2, and L3 is shown in Fig.~\ref{fig:fig2}B, and their geometrical definitions are illustrated in Fig.~\ref{fig:fig2}C. Because the vacancy framework introduces local asymmetry into the crystal structure, multiple interlayer Fe--Fe distances naturally emerge. In the Fe$_4$ parent compound, L1 and L2 remain well separated throughout the temperature range, reflecting the persistence of layer mismatch associated with the distorted Fe-sublattice configuration. In contrast, the Fe$_{4.2}$ sample exhibits a strikingly different evolution in which L1 and L2 progressively converge upon cooling and approach nearly identical values at low temperature.
	
	The physical origin of this convergence may be traced to the suppression of the structural rearrangement identified in Fig.~\ref{fig:fig1}C. In the Fe$_4$ parent compound, contraction of the Fe sublattice is accompanied by a rotational distortion that generates layer mismatch and results in L1 $\neq$ L2. In contrast, excess Fe suppresses both the contraction and rotational rearrangement, reducing the geometric source of layer mismatch. As a consequence, the Fe layers evolve toward a common interlayer geometry, driving L1 and L2 toward the same value.
	
	This behavior differs fundamentally from the in-plane evolution shown in Fig.~\ref{fig:fig1}. Whereas D1, D2, and D3 remain differentiated within the Fe$_{4.2}$ host lattice, L1 and L2 evolve toward a common value. Thus, excess Fe simultaneously preserves in-plane structural differentiation while promoting coherence along the c-axis direction. The host lattice therefore becomes increasingly differentiated within the ab plane while becoming increasingly coherent between neighboring Fe layers.
	
	Figure~\ref{fig:fig2}D provides a real-space visualization of this evolution. In the Fe$_4$ parent compound, L1 remains different from L2 owing to the distorted Fe-sublattice configuration, causing the Fe layers to deviate from an ideal planar geometry. In contrast, the Fe$_{4.2}$ sample approaches the condition L1 = L2, yielding a nearly ideal 180$^\circ$ Fe-layer configuration. We attribute this evolution to the suppression of the in-plane contraction and rotational rearrangement identified in Fig.~\ref{fig:fig1}C. By reducing these distortions, excess Fe simultaneously suppresses the associated out-of-plane response, allowing the Fe layers to evolve toward a coherent planar structure.
	
	An additional and important observation is that the d$_2$ shoulder associated with the (002) reflection gradually disappears upon further cooling, whereas the in-plane d$_2$ reflections remain clearly visible. Such anisotropic evolution is difficult to reconcile with a conventional second-phase picture, in which all d$_2$ reflections would be expected to exhibit similar temperature dependence. Instead, the present results suggest that the in-plane structural differentiation responsible for the d$_2$ features is preserved, while the layer mismatch along the c axis is progressively eliminated through the development of Fe-layer coherence.
	
	Taken together, these results show that excess Fe drives the host lattice toward a state characterized by in-plane structural differentiation and out-of-plane Fe-layer coherence. While differentiated local environments are preserved within the Fe plane, the interlayer geometry progressively evolves toward a common low-temperature configuration. The resulting structure combines two distinct in-plane environments with an increasingly coherent arrangement of neighboring Fe layers along the $c$ axis. These two characteristics constitute the defining structural features of the superconducting state and provide the structural foundation for understanding the diffraction response examined below.
	
	\begin{figure*}[t]
		\centering
		\includegraphics[width=0.8\textwidth]{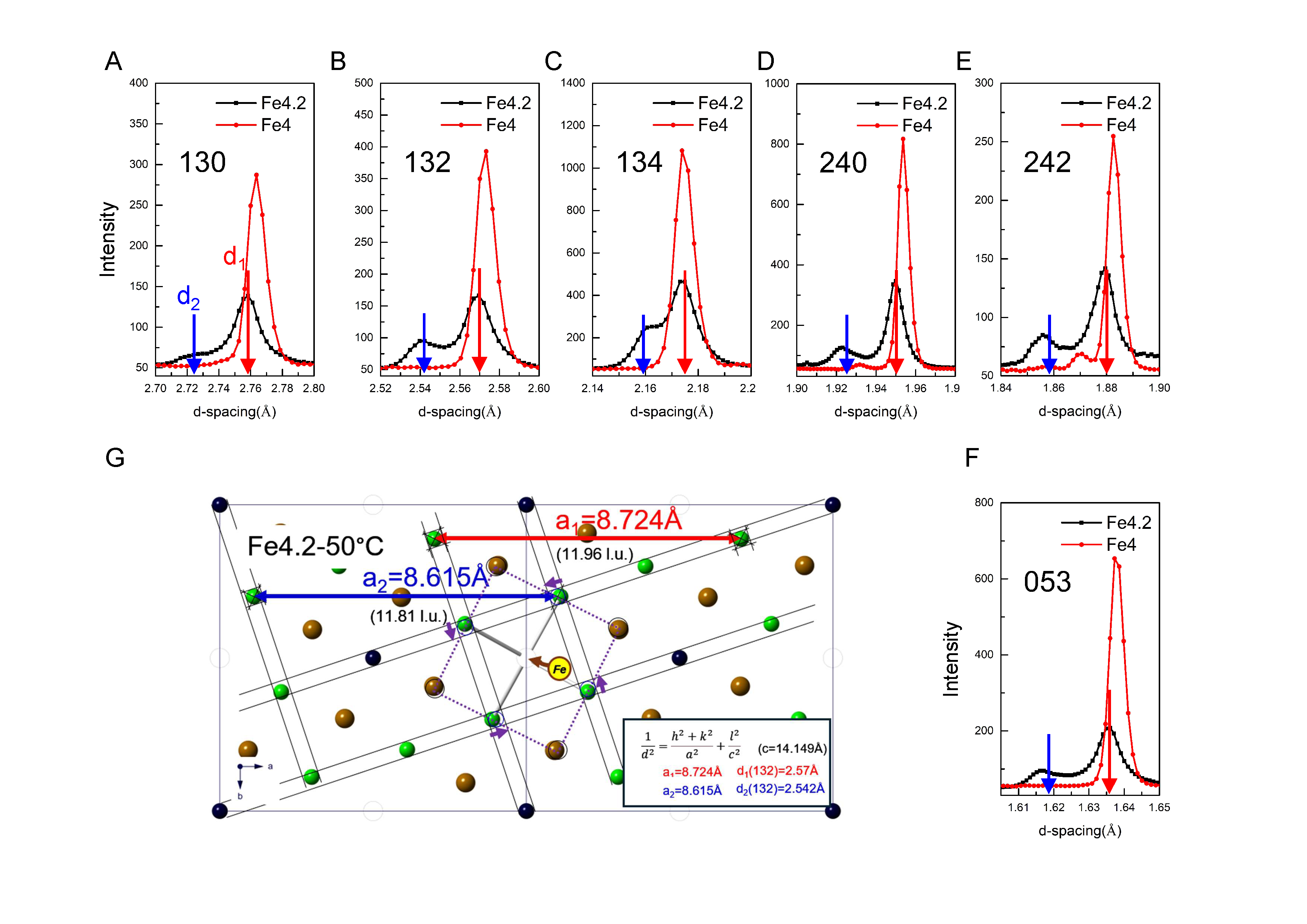}
		\caption{\textbf{Diffraction verification of the $d_2$ diffraction family.}
			(A--F) Diffraction profiles of the complete experimentally observable $d_2$ diffraction family in superconducting Fe$_{4.2}$ together with the corresponding reflections in nonsuperconducting Fe$_4$. Red (blue) arrows indicate the calculated $d_1$ ($d_2$) positions obtained using two characteristic in-plane spacings ($a_1$ and $a_2$) and a single host-lattice $c$ parameter. (G) Real-space illustration derived from the refined Se atomic positions. A representative diffraction calculation based on the two-$ab$-plus-one-$c$ framework is shown in the lower-right panel.}
		\label{fig:fig3}
	\end{figure*}
	
	To further examine the structural consequences of host-lattice differentiation, we investigate the complete experimentally observable $d_2$ diffraction family in the superconducting Fe$_{4.2}$ sample. Representative diffraction profiles are shown in Fig.~3(A--F). Additional diffraction intensity develops on the low-$d$-spacing side of the principal host-lattice reflections, producing the characteristic $d_2$ features. Rather than appearing as isolated diffraction peaks, the diffraction intensity evolves continuously from $d_1$ to $d_2$ throughout the observable diffraction family, including the (130), (132), (134), (240), (242), and (053) reflections.
	
	The significance of this behavior becomes apparent when compared with the vacancy-superlattice reflection (at near 1.932 \AA) shown in Fig.~3(D). In the Fe$_4$ parent compound, the vacancy-superlattice peak is clearly separated from the neighboring host-lattice reflection by a distinct intensity minimum, demonstrating that the instrumental resolution is sufficient to resolve closely spaced diffraction peaks. In contrast, the diffraction intensity associated with the $d_2$ features remains continuously connected to the corresponding $d_1$ reflections throughout the entire diffraction family. Even for the most clearly resolved example, the (053) reflection shown in Fig.~3(F) evolves smoothly between $d_1$ and $d_2$ without forming fully separated peaks. This behavior indicates that the observed diffraction features arise from closely related local structural environments within the host lattice.
	
	\begin{figure*}[t!]
		\centering
		\includegraphics[width=0.8\textwidth]{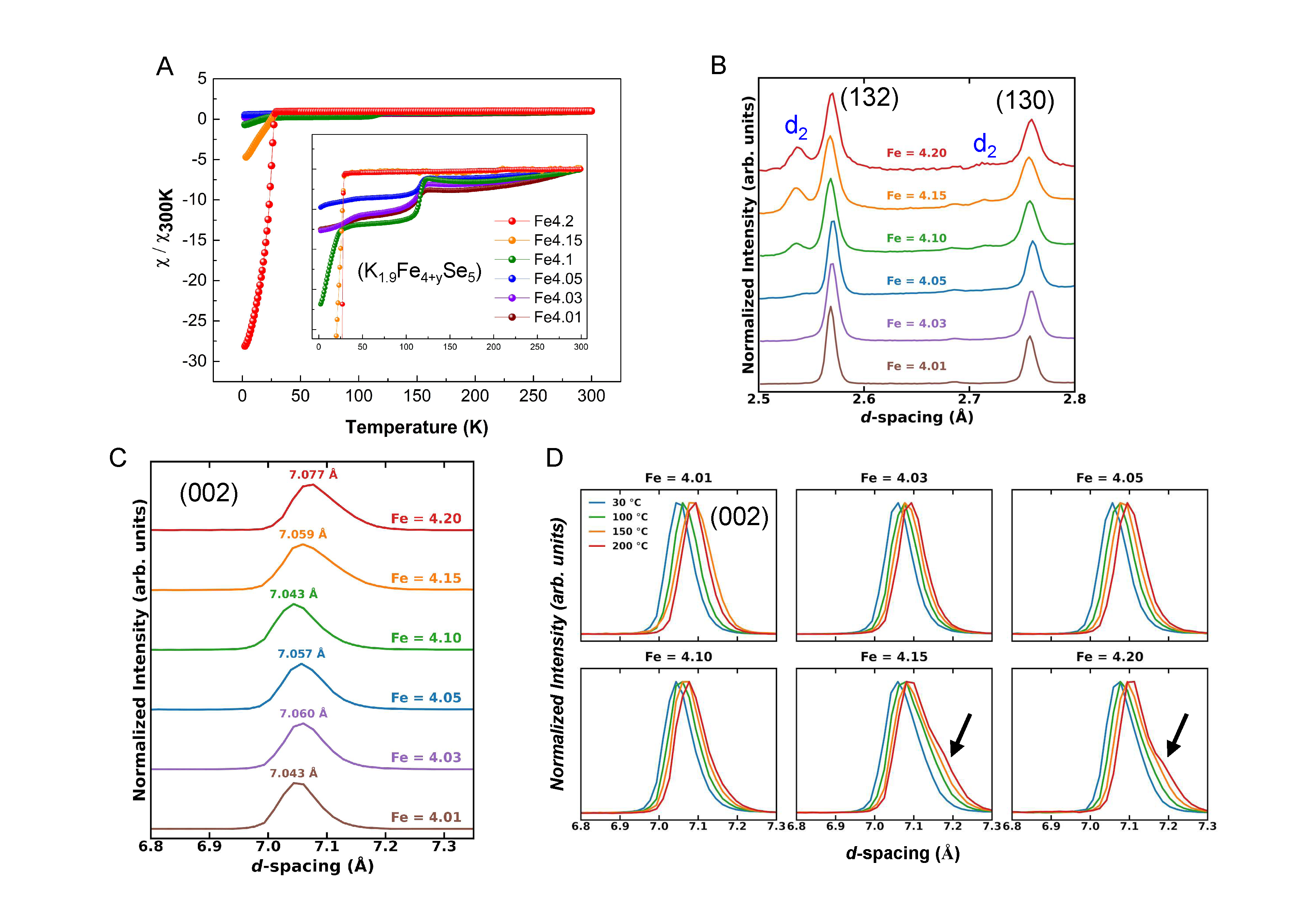}
		\caption{\textbf{Composition dependence of superconductivity and structural evolution in K$_{1.9}$Fe$_{4+y}$Se$_5$.}
			(A) Temperature-dependent magnetic susceptibility for samples with different Fe contents. (B) Evolution of the (130) reflection as a function of Fe content. (C) Composition dependence of the (002) reflection. (D) Temperature evolution of the (002) reflection for selected Fe concentrations, showing the reproducible diffraction response near $\sim200^\circ$C in superconducting samples.}
		\label{fig:fig4}
	\end{figure*}
	
	The results of Figs.~1 and 2 provide two important structural constraints for understanding these observations. First, excess Fe generates differentiated in-plane environments within the host lattice, giving rise to two characteristic in-plane spacings. Second, the out-of-plane differentiation becomes progressively suppressed upon cooling, as evidenced by the convergence of $L_1$ and $L_2$ together with the gradual disappearance of the additional contribution to the (002) reflection. The low-temperature superconducting structure therefore approaches a state characterized by two distinct in-plane spacings together with a nearly unique $c$-axis spacing. Together, these features establish the in-plane differentiation and out-of-plane coherence that characterize the superconducting state.
	
	Motivated by these experimentally established structural characteristics, we systematically examine whether all experimentally observable $d_1$--$d_2$ reflection pairs satisfy the diffraction conditions of such a lattice. The real-space model used for this analysis is shown in Fig.~3(G) and is constructed directly from the refined Se atomic positions of the host lattice. As illustrated in the local structural model in Fig.~3(G), the spatial symmetries exhibited by the Fe and Se sublattices in the $ab$ plane show subtle yet notable differences upon vacancy occupation, hinting at a differentiated local response that will be further addressed in Fig.~5. Structural differentiation generates two characteristic in-plane spacings, $a_1$ and $a_2$, while the low-temperature out-of-plane structure is represented by a nearly common host-lattice $c$ parameter. Using these experimentally determined structural parameters, the corresponding diffraction spacings can be calculated directly from the tetragonal diffraction geometry and are indicated in Fig.~3(A--F).

	The calculated spacings account for all experimentally observable $d_1$--$d_2$ reflection pairs without introducing reflection-dependent fitting parameters. Once two characteristic in-plane spacings coexist within a structure characterized by a nearly common low-temperature $c$ axis, every reflection containing an in-plane component naturally develops a corresponding second $d$ spacing. The complete $d_2$ diffraction family therefore emerges as a direct consequence of the low-temperature host-lattice geometry established in Figs.~1 and 2.
	
	Minor deviations between the calculated and experimental $d_2$ positions are expected because the present construction represents the differentiated host lattice using only two characteristic in-plane spacings and a nearly
	common c parameter. Moreover, the observed $d_2$ features appear as broadened diffraction intensity rather than fully resolved Bragg reflections, suggesting a distribution of local structural environments beyond this simplified geometric description. The continuous diffraction intensity observed between the $d_1$ and $d_2$ regions further indicates that these environments evolve continuously rather than forming perfectly separated crystallographic components. Nevertheless, the ability of the model to capture the complete experimentally observable $d_2$ diffraction family demonstrates that the essential structural information is already contained within the differentiated host lattice.
	
	The agreement obtained from the Se-based reconstruction is particularly significant when considered together with the Fe-based reconstruction shown in Fig.~1(D). Because the observed diffraction family can be accounted for using either refined Fe or Se atomic positions, the structural information responsible for the additional diffraction features is encoded in the host lattice itself rather than in a specific atomic sublattice.
	
	This interpretation is also consistent with previously reported $\sqrt{2}\times\sqrt{2}$ superlattice signatures in superconducting alkali-metal iron selenides \cite{41}. Within the framework of Fig.~3(G), partial occupation of vacancy-related sites by excess Fe locally modifies the surrounding Fe--Se coordination environment and naturally generates a $\sqrt{2}\times\sqrt{2}$ in-plane periodicity. The observation of a $\sqrt{2}\times\sqrt{2}$ superlattice signature therefore supports the local structural motif underlying the present model.
	
	Figure~2 shows that the out-of-plane differentiation becomes progressively suppressed upon cooling, while Fig.~3 demonstrates that the complete experimentally observable $d_2$ diffraction family is captured within a structural framework characterized by two characteristic in-plane spacings and a nearly common low-temperature $c$ parameter. The coexistence of multiple characteristic lattice spacings within a differentiated host lattice naturally generates diffraction signatures that resemble multiple structural components in reciprocal space, providing a natural explanation for the apparent success of conventional $I4/m+I4/mmm$ refinement procedures. An important distinction, however, is that the characteristic in-plane spacings in the present framework evolve continuously with Fe occupancy and host-lattice differentiation. The resulting diffraction response therefore forms a continuous $d_1$--$d_2$ intensity distribution, consistent with the experimentally observed peak profiles throughout the complete $d_2$ family.

	
	\begin{figure*}[t]
		\centering
		\includegraphics[width=0.8\textwidth]{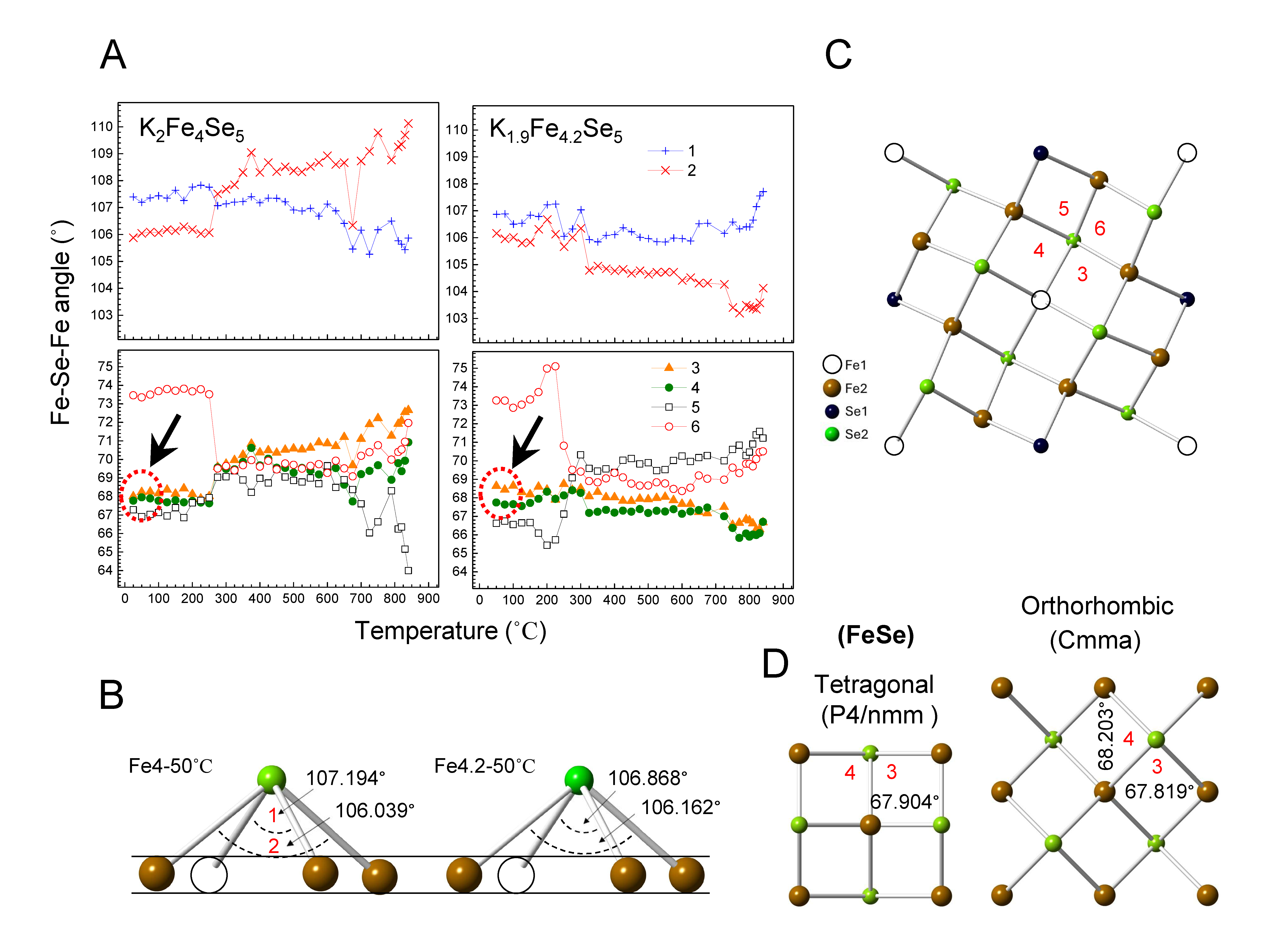}
		\caption{\textbf{Evolution of the Fe--Se--Fe pyramidal geometry and its relationship to structural differentiation.} (A) Temperature dependence of the in-plane Fe--Se--Fe bond angles. The superconducting Fe$_{4.2}$ sample exhibits a persistent bond-angle splitting, reflecting the development of differentiated local environments within the Fe plane. (B) Temperature dependence of the out-of-plane Fe--Se--Fe geometry, showing progressive convergence toward a nearly 180$^\circ$ configuration upon cooling. (C) Top-view structural diagram defining the Fe--Se--Fe bond angles (Angles 3--6) used in the analysis. (D) Fe--Se--Fe bond-angle splitting associated with the tetragonal-to-orthorhombic structural transition in FeSe, calculated from the crystallographic parameters reported by Margadonna \textit{et al.} \cite{11}.}
		\label{fig:fig5}
	\end{figure*}
	

	To further investigate the relationship between superconductivity and host-lattice evolution, we examined a series of K$_{1.9}$Fe$_{4+y}$Se$_5$ samples with different Fe concentrations. As shown in Fig.~4(A), the superconducting volume fraction increases systematically with increasing Fe content and becomes substantial when the Fe concentration exceeds approximately 4.1. The corresponding structural evolution is shown in Figs.~4(B) and 4(C). With increasing Fe content, additional diffraction intensity associated with the $d_2$ feature gradually develops. More importantly, the principal host-lattice reflections themselves undergo systematic changes. The (130) and (002) reflections both exhibit noticeable broadening, indicating increasing structural differentiation within the host lattice. At the same time, the peak positions evolve with composition. In particular, the refined $c$-axis parameter exhibits a systematic composition dependence across the superconducting region. Because these changes occur in the dominant host-lattice reflections rather than solely in the additional diffraction component, they demonstrate that superconductivity is accompanied by continuous structural evolution of the host lattice itself.
	
	More direct evidence is provided by the temperature evolution of the (002) reflection shown in Fig.~4(D). For superconducting compositions, an additional shoulder develops reproducibly near $\sim200^\circ$C during cooling. The same diffraction response is observed throughout the superconducting composition range and becomes increasingly evident as the superconducting volume fraction increases. In contrast, this behavior is absent in the nonsuperconducting compositions, indicating a close connection between the structural evolution near $\sim200^\circ$C and the emergence of superconductivity.
	
	An important feature of this diffraction response is its subsequent evolution upon further cooling. Rather than developing into a well-separated diffraction peak, the additional shoulder gradually weakens and merges back into the principal (002) reflection. Such behavior is difficult to reconcile with a conventional phase-separation picture, in which the diffraction signature of an independent structural phase would be expected to remain or become more pronounced at lower temperature. Instead, the observations suggest that the diffraction response near $\sim200^\circ$C represents an intermediate structural state along the evolution toward the low-temperature superconducting structure. The systematic appearance of this behavior exclusively in superconducting samples establishes the structural evolution near $\sim200^\circ$C as a robust and reproducible characteristic of superconducting K$_{1.9}$Fe$_{4+y}$Se$_5$.
	
    To understand the origin of this local lattice differentiation, we focus on the Se anion framework and its interplay with the interstitial Fe ions in Fig.~5. This dichotomy originates from the distinct local responses of the constituent sublattices: the Se anion framework provides a resilient tetrahedral matrix that undergoes coordinated displacive modulations in response to local potential changes induced by excess Fe, thereby sustaining a well-defined $\sqrt{2}\times\sqrt{2}$ in-plane superlattice periodicity. In contrast, the interstitial Fe ions are subject to local compositional fluctuations and site disorder, directly driving local distortions and angular discrepancies (such as the inequality between Angles 3 and 4). This coexistence—where the Se framework maintains the underlying $\sqrt{2}\times\sqrt{2}$ symmetry while Fe introduces local angular and spatial disorder—naturally reconciles the observed continuous $d_1\text{--}d_2$ diffraction profiles without invoking macroscopic phase separation.
    
    To further explore the geometric origin of the structural evolution identified above, we examine the temperature dependence of the local Fe--Se--Fe pyramidal geometry. Figure~5(A) summarizes the evolution of the relevant Fe--Se--Fe bond angles in the superconducting $\mathrm{Fe}_{4.2}$ and nonsuperconducting $\mathrm{Fe}_{4.0}$ samples. The upper panel corresponds to Angles 1 and 2 shown in Fig.~5(B), whereas the lower panel corresponds to Angles 3 and 4 illustrated in Fig.~5(C).
    
    A striking contrast is observed between the superconducting and nonsuperconducting structural pathways. For Angles 1 and 2, the nonsuperconducting $\mathrm{Fe}_{4.0}$ sample retains a persistent angular splitting throughout the measured temperature range, indicating that the out-of-plane structural mismatch remains largely preserved. In contrast, the corresponding angles in the superconducting $\mathrm{Fe}_{4.2}$ sample progressively converge upon cooling, consistent with the enhanced interlayer coherence inferred from the $L_1\text{--}L_2$ evolution shown in Fig.~2.
	
	The opposite behavior is observed for Angles 3 and 4. The nonsuperconducting structure maintains nearly equivalent bond angles, consistent with the higher in-plane symmetry of the vacancy-ordered $I4/m$ framework. In contrast, the superconducting Fe$_{4.2}$ sample exhibits persistent angular splitting, indicating a reduction of local in-plane symmetry and providing a natural structural origin for the broadened and split diffraction features observed in superconducting samples.
	
	Viewed together, the bond-angle evolution reveals nearly opposite symmetry-evolution pathways. The nonsuperconducting state preserves high in-plane symmetry while retaining substantial out-of-plane differentiation. In contrast, the superconducting state progressively suppresses the out-of-plane differentiation while preserving in-plane structural differentiation. Superconductivity therefore appears to be associated not with structural homogenization, but with the coexistence of enhanced out-of-plane coherence and persistent in-plane differentiation.
	
	For comparison, Fig.~5(D) reproduces the bond-angle evolution associated with the tetragonal-to-orthorhombic transition in FeSe. In the tetragonal phase, Angles 3 and 4 are symmetry-equivalent, whereas the orthorhombic phase develops a finite angular splitting corresponding to inequivalent in-plane Fe environments. Interestingly, a closely related behavior is observed in the present system: the nonsuperconducting K$_{2-x}$Fe$_{4+y}$Se$_5$ structure retains nearly equivalent Angles 3 and 4, whereas the superconducting structure develops and preserves substantial angular splitting. Thus, despite their fundamentally different crystallographic frameworks, both systems associate superconductivity with a reduction of in-plane symmetry. Moreover, orthorhombic FeSe retains a highly coherent out-of-plane geometry, closely resembling the enhanced interlayer coherence observed in the superconducting K$_{2-x}$Fe$_{4+y}$Se$_5$ samples.
	
	An interesting connection can also be made with the scanning tunneling microscopy study of Li \textit{et al.}, which identified a superconducting region characterized by a $\sqrt{2}\times\sqrt{5}$ ordering pattern coexisting with a nonsuperconducting $\sqrt{2}\times\sqrt{2}$ region \cite{21}. Remarkably, the superconducting region occupied approximately five times the area of the nonsuperconducting region, indicating that superconductivity is associated with the dominant structural environment rather than a minor impurity phase. Within the framework proposed in Fig.~5(C), occupation of vacancy-related sites by excess Fe generates a local $\sqrt{2}\times\sqrt{2}$ environment embedded within the surrounding $\sqrt{5}\times\sqrt{5}$ vacancy-ordered host lattice. The coexistence of locally reconstructed and vacancy-ordered environments therefore bears a strong structural resemblance to the $\sqrt{2}\times\sqrt{5}$ superconducting state reported by Li \textit{et al.}. From this perspective, the superconducting ordering signatures may be understood as another manifestation of the host-lattice differentiation identified in the present work. A recent study likewise associated superconductivity in K$_x$Fe$_{2-y}$Se$_2$ with regions retaining nearly intact FeSe layers, further emphasizing the importance of the underlying FeSe-based structural framework \cite{42}. Crucially, this nanoscale coexistence of locally reconstructed and vacancy-ordered domains provides a direct real-space origin for the continuous intensity distribution observed in macroscale XRD profiles, bridging local electronic inhomogeneities with global structural evolution.

\section{Conclusion}
Using high-resolution synchrotron X-ray diffraction, we have elucidated the structural evolution of K$_{2-x}$Fe$_{4+y}$Se$_5$ as a function of Fe content, temperature, and thermal history. Excess Fe suppresses the vacancy-ordering structural response of the parent lattice, generating in-plane structural differentiation prior to the emergence of additional $d_2$ diffraction features. Real-space reconstructions and geometric analysis demonstrate that the complete $d_2$ diffraction family is naturally captured within a unified host-lattice framework characterized by two characteristic in-plane spacings and a nearly common low-temperature $c$ parameter. This accounts for the apparent success of conventional two-phase refinements as a natural consequence of continuous host-lattice differentiation.

Furthermore, temperature-dependent measurements reveal distinct structural pathways in superconducting samples: while in-plane differentiation is preserved, the interlayer geometry simultaneously develops enhanced Fe-layer coherence along the $c$ axis upon cooling. Analysis of the Fe--Se--Fe pyramidal geometry further shows that superconductivity is associated with reduced in-plane symmetry and enhanced out-of-plane coherence. These findings establish host-lattice evolution as the common structural origin linking local differentiation, the d$_2$ diffraction family, and superconductivity in K$_{2-x}$Fe$_{4+y}$Se$_5$. The close correspondence between the host-lattice differentiation identified here and the symmetry redistribution associated with superconductivity in FeSe suggests that these phenomena may represent structurally analogous manifestations of a common superconducting pathway in FeSe-based systems.

	
	\begin{acknowledgments}
		We acknowledge the National Synchrotron Radiation Research Center (NSRRC) in Taiwan for providing the beamtime. We are particularly grateful to Hwo-Shuenn Sheu and Jey-Jau Lee for their invaluable technical support with the high-resolution synchrotron X-ray diffraction measurements.
	\end{acknowledgments}


\end{document}